\documentclass[twocolumn,prl,aps, 10pt]{revtex4-1}
\usepackage{amsmath, amsthm,amsfonts,graphicx,verbatim, color, braket, dsfont, amssymb, times, hyperref}
\usepackage{geometry}                		
\usepackage{graphicx}				

\newcommand{\gi}{  \lambda_{\textrm{I}}}
\newcommand{\gr}{  \lambda_{\textrm{R}}}
\newcommand{\gk}{  \lambda_{\textrm{KM}}}

\begin{document}

\title{The gate-tunable  strong and fragile topology of multilayer-graphene on a transition metal dichalcogenide}

\author{Michael P. Zaletel}
\affiliation{Department of Physics, University of California, Berkeley, CA 94720, USA}
\author{Jun Yong Khoo}
\affiliation{Department of Physics, Massachusetts Institute of Technology, Cambridge, Massachusetts 02139, USA}

\begin{abstract}

	We analyze the phase diagram of multilayer-graphene sandwiched between identical transition metal dichalcogenides.
Recently realized in all van-der-Wall heterostructures, these sandwiches induce sizable (1-15 meV)  spin orbit coupling in the graphene, offering a way to engineer topological band-structures in a pristine and gate-tunable platform.
We find a rich phase diagram that depends on the number of layers $N$ and the gate-tunable perpendicular electric field. For $N > 1$ and odd, the system is a strong 2D topological insulator  with a gap equal to the strength of proximity-induced Ising spin-orbit coupling, which reverts to a trivial phase at moderate electric fields.
For $N$-even, the low energy bands exhibit  a recently proposed  form of ``fragile'' crystalline topology, as well as electric-field tuned symmetry-protected phase transitions between distinct atomic insulators.
Hence AB-stacked bilayer and ABC-stacked trilayer graphene  are predicted to provide controllable experimental realizations of fragile and strong topology.
	
\end{abstract}

\maketitle

	Spin-orbit coupling (SOC) is an essential ingredient for  realizing  2D topological insulators (TI)  featuring gapless edge states protected by time-reversal symmetry.\cite{hasan2010colloquium, qi2011topological}
In principle, the SOC coupling intrinsic to graphene (the Kane-Mele mass) is already sufficient to realize a strong TI,\cite{KaneQSH} but in practice  it is extremely weak.\cite{Min2006, YaoSO2006, sichau2017intrinsic}
This has led to an effort to ``proximitize'' SOC in graphene 
\cite{avsar2014spin, Wang2016Origin, yang2016, Volkl2017, Yang2017, Wakamura2018, wang2015,  kaloni2014quantum, Gmitra2015, Gmitra2016, Cysne2018,  Khoo2017, khoo2018tunable}
 by placing it on an insulating transition metal dichalcogenide (TMD) substrate.
At low energies, DFT calculations predict that the dominant effect of the TMD is to induce various types of translation-invariant SOC in the proximate graphene layer, but the Kane-Mele type remains small.\cite{kaloni2014quantum, Gmitra2015, wang2015, yang2016, Gmitra2016, Cysne2018}
Experiments  have indeed found evidence for some form of proximitized SOC in graphene, on the order of 10meV,\cite{avsar2014spin, wang2015, Wang2016Origin, yang2016, Volkl2017, Yang2017, Wakamura2018} but its precise nature, and  a complete understanding of the topological phases it might   enable, remained unclear.

	Recent experiments on bilayer graphene (BLG) in contact with a  TMD on one or both sides have shed new light on this issue.\cite{Island}
By comparing thermodynamic compressibility measurements against theoretical modeling, clear evidence was found for ``Ising'' spin-orbit coupling of magnitude $\gi \sim 2.5$meV (cf. Eq~\eqref{eq:mlg}).
Furthermore, in double-sided devices, these experiments find that $\gi$ gaps out BLG's quadratic band touching\cite{mccann2013electronic} in a different manner than a perpendicular electric field or sublattice splitting.
With increasing  electric field, this spin-orbit induced gap closes, and then reopens, indicating the presence of  distinct band insulators within the phase diagram. 
The  existence of a phase transition in the  absence of any apparent symmetry breaking is the hallmark of a topological phase transition, as would occur, for example, between a  strong  2D TI and a trivial phase.

In this work we determine the precise nature of the topology enabled by the SOC, both in BLG and more generally in $N$-layers of chirally-stacked  graphene. 
For $N > 1$ and odd, we find the system is a strong TI, making ABC-trilayer / TMD sandwiches an interesting direction for future experiments.
The $N$-even case is more subtle: we show that the SOC-induced gap observed in BLG is not a strong TI,  but rather an  example of a topological crystalline insulator (TCI)   with  ``fragile'' topology.
\cite{po2016filling, Shiozaki2017, PoWatanabeFragile, bradlyn2017topological, bradlyn2018disconnected, liu2018shift, else2018fragile}
We emphasize at the outset that these TCI phases do not have protected edge states,\cite{fu2011topological} and (for a suitably generous definition of ``deform''\cite{else2018fragile})  they can be smoothly deformed to a Slater determinant of symmetric, localized Wannier orbitals,\cite{MarzariMLWF} in contrast to Chern bands and strong TIs.
What is interesting about these phases is rather \emph{where} the Wannier orbitals are localized,\cite{Zak1982, king1993theory, MarzariMLWF, PoHomotopy, Shiozaki2017, huang2017building} because the $C_{3v}$ point group symmetry pins their allowed positions, leading to quantized polarization and  multipole moments.\cite{king1993theory, MarzariMLWF, benalcazar2017electric}
Two insulators with different quantized moments  must be separated by a phase transition, which we argue explains the  transition observed in the BLG experiments.

\emph{MLG on TMD}.
Throughout, we use $\tau,  s, \sigma$ to denote valley, spin and sublattice respectively. 
The low-energy Hamiltonian of monolayer graphene on a TMD can be parameterized phenomenologically as \cite{wang2015, Gmitra2015}
\begin{multline}
H_{\textrm{MLG}} = v_F( \tau^z \sigma^x k_x + \sigma^y k_y ) +  \frac{m}{2} \sigma^z \\ +  \frac{\gi}{2}  \tau^z s^z  + \frac{\gr}{2} ( \tau^z \sigma^x s^y - \sigma^y s^x ) + \frac{\gk}{2} \tau^z \sigma^z s^z
\label{eq:mlg}
\end{multline}
The TMD induces the sublattice splitting $m$,  ``Ising'' SOC $\gi$, and  Rashba SOC $\gr$. The intrinsic SOC $\gk \sim 40 \mu$eV (Kane-Mele mass) is vanishingly small  so we neglect it.\cite{Min2006, YaoSO2006, sichau2017intrinsic}
DFT calculations estimate $m \sim 0-1, \gi \sim 1-5,  \gr \sim 1 - 15$ meV depending on the TMD.\cite{wang2015, Gmitra2015, Gmitra2016, yang2016}
While initially it was suggested MLG on a TMD might be a strong 2D TI,\cite{kaloni2014quantum, wang2015, Gmitra2016} explicit calculation shows the $\mathbb{Z}_2$ index  is trivial.\cite{yang2016}

To analyze the symmetries, note that mirrors act as $M_x = i \tau^x s^x, M_y = i  \sigma^x s^y, M_z = -i s^z$, time-reversal as $\mathcal{T} = i \tau^x  s^y$, and rotation as $C_3 = e^{i \frac{2\pi}{3}( -\tau^z \sigma^z + s^z / 2) }$.
So $m$ and $\gi$ are odd under $M_y$, while $\gr$ is odd under $M_z$, both of which are broken by the TMD. $M_x$ remains a good symmetry, so the space group is $p3m1$ (in class AII).

\begin{figure}[t]
\includegraphics[width=\columnwidth]{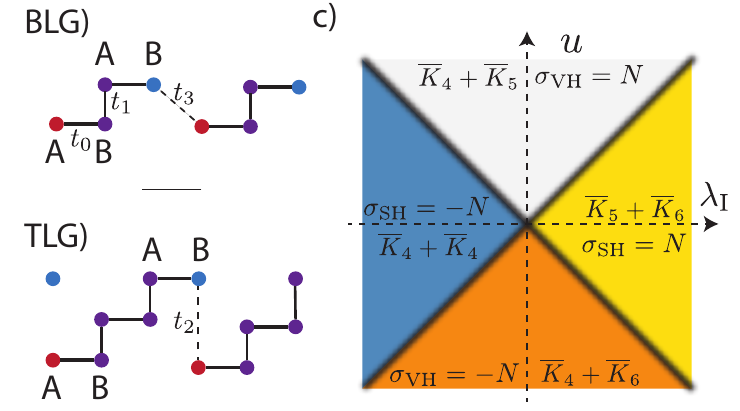}
\caption{\label{phased} Cross section of the \textbf{BLG}) and \textbf{TLG}) lattice structure.  The dominant hopping $t_0$ is intra-layer hopping (e.g. graphene), and the vertical hopping $t_1$ leaves behind two low-energy orbitals of a/b type on the bottom / top layer. \textbf{c}) Phase diagram of $N$-layer graphene with perpendicular electric field $u$ and Ising SOC $\gi$.
In the simplest analysis the four phases are distinguished by their spin and valley Hall coefficient $\sigma_{\textrm{SH} / \textrm{VH}}$.
These are not quantized once accounting for Rashba $\gr$ and lattice effects, but the $\sigma_{\mathrm{SH}}=$odd phase descends to a strong 2D $\mathbb{Z}_2$ TI. Hence TLG on a TMD is a $\mathbb{Z}_2$ TI. 
}
\end{figure}

\emph{Effective Hamiltonians for TMD  / multilayer graphene / TMD sandwiches.}
We focus on  ``chiral'' multilayers in which the $A$ sublattice of layer $\ell+1$ is stacked on top of  the $B$-sublattice of layer $\ell$. For bilayer graphene (BLG), this is the AB Bernal stacking, while for TLG this is the ABC-stacking, as shown in Fig.~\ref{phased}.
Without the TMD, the minimal effective hopping model for the multilayer is
\begin{multline}
H = v_F( \sigma^+ k_- + \sigma^- k_+)  +  t_1( \sigma^+ \ell^- + \sigma^- \ell^+  ) \\ - u \frac{\ell - N/2}{N-1}  
\end{multline}
Here $k_{\pm} = \tau^z k_x \pm i k_y$; the interlayer hopping operator is $[\ell, \ell^{\pm}]  = \pm \ell^{\pm}$;  $t_1 \approx 0.36$meV is the interlayer hybridization in the notation of Ref.~\onlinecite{jung2014accurate}; and $u$ is the potential difference between the bottom and top layer due to an electric field.
We drop further-neighbor hoppings $t_2, t_3, t_4$ of the Slonczewski-Weiss-McClure (SWM) model, but they will be included in numerical band calculations and have no impact on our conclusions.

	Following earlier approaches,\cite{Khoo2017, khoo2018tunable} we then assume the TMD couples only to the two outer layers according to Eq.~\eqref{eq:mlg}. Because the Rashba coupling is odd under $M_z$, if the same TMD is used for the top / bottom substrate we expect equal and opposite couplings on the top / bottom layer, $\gr = \gr^b = -\gr^t$. For $\gi, m$, however, there are two possibilities.
If the stack has a 3D inversion symmetry $\mathcal{I} = M_x M_y M_z$, then  $\gi = \gi^b = -\gi^t$. But if the top TMD is then rotated by 180$^\circ$, we will have $\gi^b= \gi^t$ (and similarly for $m$), a case discussed in Ref.~\onlinecite{khoo2018tunable}.
A priori both configurations may be metastable (as well as mis-aligned intermediate cases), but in the  experiments of Island et al.~\cite{Island} most samples are consistent with  inversion symmetry, so here we restrict to this case and refer to the Appendix for the general one. For any finite $u$, $\mathcal{I}$ is broken and the wallpaper group is again $p3m1$.
 
The hybridization $t_1$ gaps out most of the orbitals, but leaves orbitals $\ket{ \ell = 1, A}$ and $\ket{\ell = N, B}$ unpaired. We follow the standard procedure for integrating out the hybridized orbitals\cite{mccann2013electronic, zhang2010band, khoo2018tunable} to obtain an effective Hamiltonian for the two low-energy bands to lowest order in $t_1^{-1}$,\footnote{A different limit, $\gr \sim t_1$ is analyzed in Ref.\onlinecite{qiao2011two} but it appears this can't be reached on TMDs.} as detailed in the Appendix.
We obtain
\begin{align}
\label{eq:heff}
H_{\textrm{eff}} &= \left(\begin{array}{cc} \frac{u}{2} + \frac{\gi}{2} \tau^z s^z &  (v_F k_-)^{N}  / t_1^{N-1}  \\  (v_F k_+)^{N} / t_1^{N-1} & -\frac{u}{2} - \frac{\gi}{2} \tau^z s^z  \end{array}\right) + \cdots \\
&= H_{\textrm{NLG}} + \frac{u}{2} \sigma^z +  \frac{\gi}{2}  \tau^z \sigma^z s^z + \cdots
\end{align}
Here $\cdots$ denotes terms of order $\mathcal{O}(\frac{g^2}{t_1^2})$, where $g$ are small parameters in the model such as $\gr$ and the neglected SWM terms.
In particular, due to a cancellation between $\gr^t = -\gr^{b}$,
the Rashba coupling first appears  in the form $\gr \frac{g^2}{t_1^2}$, so is highly suppressed.

The  kinetic part $H_{\textrm{NLG}}$ is the well-known chiral band touching with dispersion $\epsilon(k) = \pm (v_F k)^N / t_1^{N-1}$, while, fortuitously, the field $u$ and Ising SOC $\gi$ are converted into the trivial mass $m$ and Kane-Mele mass $\gk$ of the monolayer case respectively.
To analyze the resulting phase diagram, we note that when a \emph{single} band-touching with chirality $N$ is gapped out by a mass ``$\Delta$'', the conductance / valence bands carry Chern-number $C = \pm \textrm{sign}(\Delta) \tau^z N / 2$ respectively.
The   $\tau^z$  dependence arises because the chirality of $k_{\pm}$ reverses with valley.
If we temporarily pretend $\tau^z, s^z$ are \emph{exactly} conserved and add up the contributions across $\tau, s$, we then obtain the phase diagram shown in Fig.~\ref{phased}c), containing valley-Hall ($|u| > |\gi|$) \cite{khoo2018tunable} and quantum spin-Hall (QSH) phases ($|\gi| > |u|$)  with Hall coefficients $\sigma_{\textrm{SH}/\textrm{VH}} = \pm N$.

However, while the effective model has continuous symmetries generated by $s^z, \tau^z$, microscopically there are only the discrete translations and spin-orbit coupled $C_{3v}$  point group.
Indeed, $s^z$ is weakly broken by $\gr$ at higher order in $t_1^{-1}$.
Consequently the spin and valley Hall coefficients, while approximate, are not strictly quantized.
Nevertheless, tight-binding simulations of the full band model including these effects and further SWM hoppings confirm there is a  semi-metallic transition between these phases for $|u| \sim |\gi|$, both for $N=2$ and $N=3$.
(Though we note in BLG the transition is mediated by an narrow intervening compensated semimetal \emph{phase}, which arises because the trigonal warping $t_3$ splits the quadratic band touchings into  $4 = 3 + |-1|$ Dirac cones, Fig.~\ref{wyckoff}.)
This suggests there is a robust topological distinction between them, which we now analyze given the actual $p3m1$ symmetry. 

\emph{$N$-odd: strong $\mathbb{Z}_2$ TI}.
\begin{figure}[t]\label{edge}
\includegraphics[width=\columnwidth]{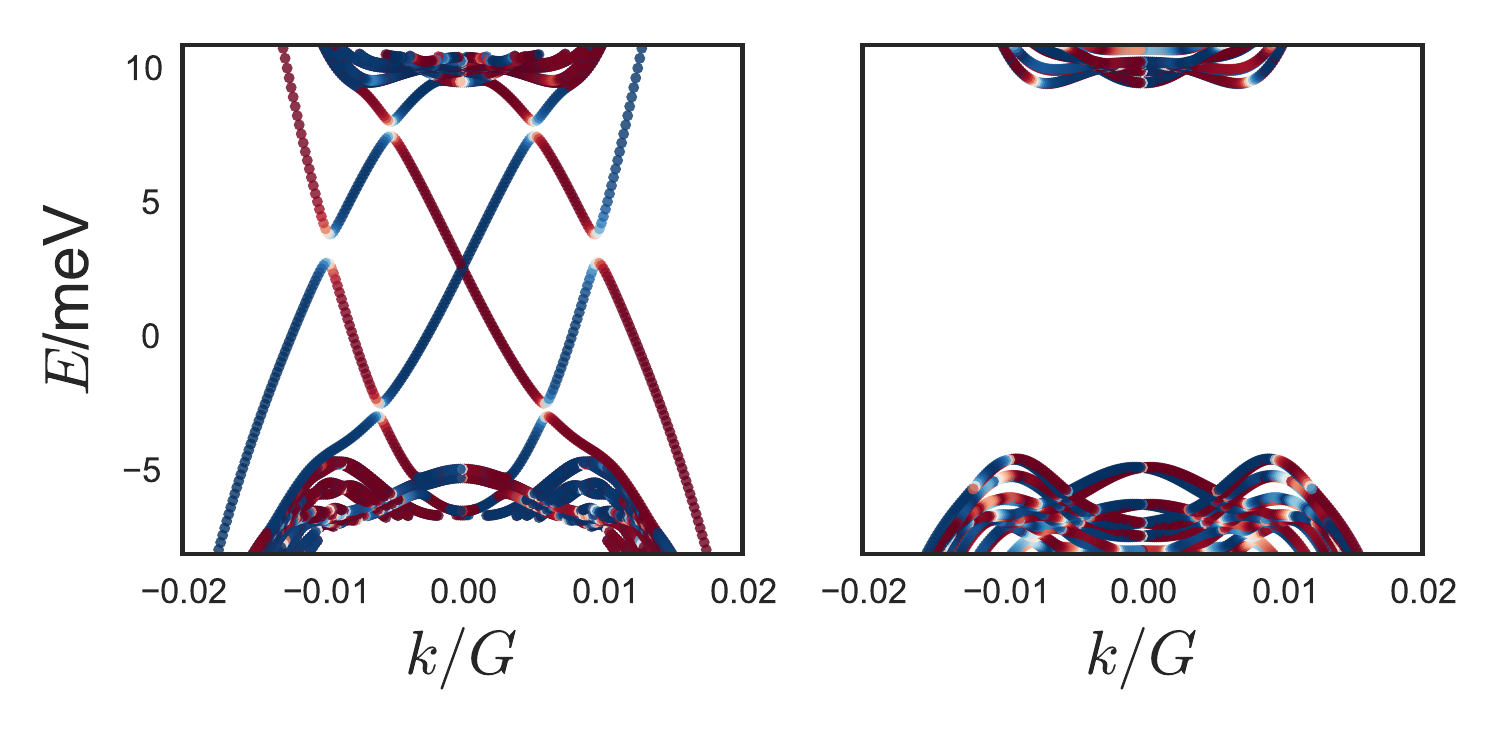}
\caption{ Edge spectrum of TLG in TMD, for \textbf{a}) topological ($u = 0$) and  \textbf{b}) trivial ($u = 30$meV) values of the electric field.
We diagonalize an arm-chair edge strip of width $W = 500 \sqrt{3} a$ for the lowest 32 states.
Color denotes the spin $\langle s^z \rangle$, and data is only shown for states with position localized to $\langle \hat{y} \rangle < \frac{3}{4} W$.
In \textbf{a}) we observe three spin-filtered sub-gap crossings, two of which are gapped out by the Rashba coupling $\gr$ and one of which (at $k=0$) is topologically protected by the $\mathbb{Z}_2$ TI index. In the trivial regime \textbf{b}), there are no sub-gap states (the zig-zag edge does have subgap states, but no exactly degenerate crossings).
Tight binding parameters are $t_0 = -2.6, t_1 = 0.36, t_2 = -0.01, t_3 = 0.28, t_4 = 0.14$ (in eV) taken from Ref.~\onlinecite{jung2014accurate}. We implement SOC  using the minimal hoppings required to reproduce Eq.~\eqref{eq:mlg}, with $\gi = 15$meV and $\gr = 25$meV, which are somewhat larger than their likely values due to computational constraints.
}
\end{figure}
For $N$-odd, $|\gi| > |u|$, the spin-Hall conductance $\sigma_{\textrm{SH}} = \pm N$ is odd.
While the weak SOC terms which break $s^z$-conservation break the $\mathbb{Z}$ QSH classification, any phase with odd spin-Hall conductance is automatically a $\mathbb{Z}_2$ TI.\cite{kane2005z}
Hence we predict the ABC trilayer-graphene / TMD sandwich at $u=0$ is a strong TI.
The $|u| > |\gi|$ phase is topologically trivial (it is adiabatically connected to the sub-lattice polarized phase), so by using a double-gate to control $u$ we obtain a gate-tunable TI.
	
	The arm-chair edge spectrum of TLG in the $u = 0$ phase is shown in Fig.~\ref{edge}a), including all the further-neighbor SWM hoppings.
As expected of a TI, we find spin-filtered subgap states with a crossing at $k=0$ protected by time-reversal.
In the absence of $\gr$ we verified there are three gapless crossings, consistent with the $\sigma_{\textrm{SH}}=3$ QSH effect. For $u > \gr$, we find an edge spectrum which is fully gapped.

\emph{$N$-even: crystalline symmetry-distinct atomic insulators and symmetry-protected phase transitions.}
The above analysis implies that $N$-even  has trivial strong-topology, but since the $|u| \sim |\gi|$ transition is robust, this suggests there is a ``crystalline topological'' distinction between the  phases which we now explicate.

For $u \gg | \gi |$, the two low-energy electrons per unit cell  both localize on orbital $A$, forming a local Kramer's pair. Likewise for $u \ll - |\gi|$, the doublet localizes on the $B$ sublattice. 
We refer to states which can be adiabatically deformed to a set of fully-filled symmetric, localized Wannier orbitals \cite{MarzariMLWF} as  ``atomic insulators,'' (AIs, or ``band representations'' in Ref.~\onlinecite{Zak1982, bradlyn2017topological}) which would seem to be the very model of trivial phases.  
However the crystal symmetry adds a new twist, because the filled orbitals of these two phases are localized at distinct  high-symmetry sites (Wyckoff positions ``a, b'' in Fig.~\ref{wyckoff}) pinned by the $C_{3v}$ symmetry.
Consequently there is no way to continuously pass between them without encountering a phase transition.
A generic phase transition between two states of the same symmetry is the essential feature of a topological phase transition,  which arises here because there is a symmetry-protected  \emph{difference} between the two AIs.

There is a third  high-symmetry point, the center of the hexagons ``c.''
While there is no site at ``c'' within the two-band model (the orbitals there are at a much higher energy $t_1$), the Wannier orbitals of a phase $C$ could nevertheless be delocalized around A/B orbitals \emph{surrounding} the hexagon so that they transform under the $C_{3v}$ symmetry of site c.
Some works would call this an ``obstructed AI''  because the low-energy model lacks the c-orbitals required to deform it to a simple product state.\cite{bradlyn2017topological, PoWatanabeFragile, Cano2018, bradlyn2018disconnected, else2018fragile}
We will show shortly that the $\gi > |u|$ region is such an obstructed phase.
Finally, for $\gi < - |u|$, we obtain a phase we  denote ``$F$.'' Interestingly, by itself $F$ has \emph{no} localized Wannier representation within the two-band model. However, there is nevertheless a local picture for this phase: start by filling bands $A + B$, and then make an atomic insulator of \emph{holes} at the sites of $C$, e.g. ``$F = A + B - C$,'' an example of  fragile topology.\cite{PoWatanabeFragile, bradlyn2018disconnected, else2018fragile} 
 
\begin{figure}[t]
\includegraphics[width=\columnwidth]{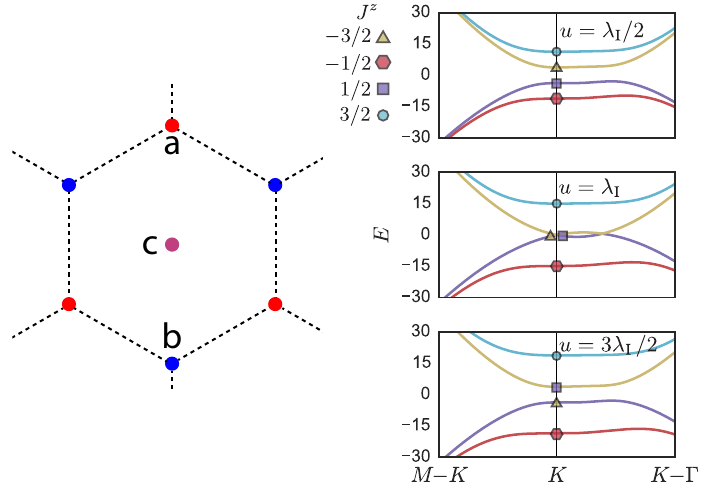}
\caption{\label{wyckoff} \textbf{Left}) Unit cell of BLG, showing the low-energy orbitals at positions ``a,'' ``b'' and the $t_1$-hybridized orbitals at ``c.''
\textbf{Right}) The $\gi, u$-driven band inversion of BLG: energy spectrum at the $K$-point for three values of electric field $u$ (band structure parameters are as in Fig.~\ref{edge}). The $J^z$ eigenvalues at the $K$-point are shown in the legend (top) and (bottom).
}
\end{figure}

	All four phases are distinct in the presence of $p3m1$ (even with interactions \cite{else2018fragile, huang2017building}), which would explain the robustness of the phase diagram.
To confirm they are obtained in our model we analyze their $k$-space representation.\cite{bacry1988symmetry, po2017symmetry, elcoro2017double}
The three  AIs $A, B, C$,  arise by filling an $S=1/2$ $p_z$-orbital ( representation $\bar{E}_1 \uparrow G(2)$ of $C_{3v}$ ) on either the a, b, or c Wyckoff positions respectively. \footnote{The three Wyckoff positions are equivalent up to a choice of origin, but if we view ``c'' as the origin it would be called position 1a in  crystallographic tables.}
Examining their representations at the high-symmetry points in $k$-space, we find that at the $\Gamma$ and $M$ point all three AIs have the same representations, $\bar{\Gamma}_6(2)$ and $\bar{M}_3 \bar{M}_4(2)$. However, at the $K$ point they are distinct : $A, B, C$ have representations $\bar{K}_4 + \bar{K}_6, \bar{K}_4 + \bar{K}_5$ and $\bar{K}_5 + \bar{K}_6$ respectively. 
To demystify the notation, we note that these representations merely encode the spin $J^z$ under $C_3$-rotations:
$\bar{K}_4$ is $J^z = -\frac{3}{2} \sim \frac{3}{2}$, $\bar{K}_5$ is $J^z = -\frac{1}{2}$ and $\bar{K}_6$ is $J^z = \frac{1}{2}$. The spins $J^z \sim J^z + 3$ are equivalent because only the $C_3$ crystal symmetry, rather than full rotations, is present.

In the two-band graphene model (Eq.~\eqref{eq:heff}), the low-energy Bloch wave-functions for $A_\uparrow, A_\downarrow, B_\uparrow, B_\downarrow$ at valley $\tau^z = 1$ have spin $J^z = \sigma^z + \frac{1}{2} s^z$.
Thus we can directly equate them with representations $\bar{K}_4, \bar{K}_6, \bar{K}_5, \bar{K}_4$ respectively.
Recalling that the band touching is gapped by $H_\Delta = \frac{u}{2} \sigma^z + \frac{\gi}{2} s^z  \sigma^z + \cdots $, we see that when  $\gi > |u|$, orbitals $A_\downarrow, B_\uparrow \sim   \bar{K}_5 + \bar{K}_6$ fill: precisely the obstructed AI $C$.
In contrast, when $\gi < -|u|$, orbitals $A_\uparrow, B_\downarrow \sim 2 \bar{K}_4$ fill, which a consultation of the Bilbao tables \cite{elcoro2017double} confirms  has \emph{no} AI representative - this is the fragile phase ``$F$,'' which is clearly just the particle-hole conjugate $F = A + B - C$.

Our analysis of the two band model neglects the $t_1$-hybridized band well below the Fermi level, which is  localized on $C$.
Adding back these filled ``core'' orbitals, the four phases are the stacked AIs $C + A$, $C + B$, $C + C$ and $ C + (A + B - C) = A+B$ respectively. 
So, strictly speaking, in the enlarged model none of the phases are obstructed or fragile. 
However, the \emph{difference} between them is still robust.

	Is the $u=0$ phase observed in BLG is $C$ or $F$, e.g., what is the sign of $\gi$?
The compressibility measurements  on double-sided devices in Ref.~\onlinecite{Island} were insensitive to the sign of $\gi$, but measurements  on a device with TMD on only one side suggest $\gi > 0$. 
It may also be that the sign is sample dependent because it depends on the TMD-BLG alignment.
	

In conclusion, the SOC found in graphene on a TMD is predicted to stabilize a strong TI in ABC-graphene, and symmetry-distinct atomic insulators separated by phase transitions in BLG. 
Within the low energy two-band model these phases  feature obstructed and fragile topology, and it would be interesting to determine whether any physical signatures of these phases, such as in flux-spin pumping \cite{liu2018shift} or novel Landau level spectra, \cite{lian2018landau}  survive once accounting for the full band structure.

\acknowledgements{
We are indebted to our collaborators Joshua Island, Leonid Levitov, Cyprian Lewandowski, and Andrea Young  for their work on BLG/TMD devices, and to conversations with Andrei Bernevig, Shenjie Huang, Hoi Chun Po and Haruki Watanabe on crystalline topology.
MZ was supported by the Director, Office of Science, Office of Basic Energy Sciences, Materials Sciences and Engineering Division of the U.S. Department of Energy under contract no. DE-AC02-05-CH11231 (van der Waals heterostructures program, KCWF16). JYK was supported by the National Science Scholarship from the Agency for Science, Technology and Research (A*STAR).}

%

\onecolumngrid
\appendix

\section{Derivation of effective Hamiltonian}
Here we derive the effective Hamiltonian in the presence of intralayer and interlayer hopping $t_0, t_1$, spin-orbit coupling  $\gr^{t/b}, \gi^{t/b}$, electric field $u$, masses $m_i$ on layer $i$, and Zeeman field $E^Z$:
\begin{align}
H &= H^{t_0} + H^{t_1}  + H^{\gr} + H^{g} \\
 H^{t_0} &= v_F ( \sigma^+ k_- + \sigma^- k_+), \quad  k_{\pm} = \tau^z k_x \pm i k_y \\
 H^{t_1}  &= t_1 (\sigma^+  \ell^-  + \sigma^-  \ell^+) \\
 H^{\gr} &=  i  \gr^{t/b} P^{\ell}_{t/b} (\sigma^+ s^- - \sigma^- s^+ ), \quad s^{\pm} \equiv ( s^x \pm i \tau^z s^y)/2 \\
 H^{g} &=  \frac{\gi^{t/b}}{2}  P^{\ell}_{t/b} \tau^z s^z + \frac{u}{N-1} (\ell - \frac{N}{2} )  + P^{\ell}_{i} \frac{m_i}{2} \sigma^z + \frac{1}{2} \mathbf{E}^Z \cdot  \vec{s}
\end{align}
where $P^\ell_i$ denotes projection into layer $\ell = i$. We use  ``$g$'' to refer collectively to the couplings in $H^g$.
Our analysis does not account for further neighbor couplings $t_2, t_3, t_4$, whose effect we comment on later.

We decompose the Hamiltonian into the $t_1$ unhybridized (1) and hybridized orbitals (2), $H = \left(\begin{array}{cc}H_{11} & H_{12} \\H_{21} & H_{22}\end{array}\right)$. 
For (22)  we decompose  $H_{22} = H_{22}^{t_0} + H_{22}^{t_1} + H_{22}^{g}$ following the decomposition above.
For (11) we have  $H_{11} = H_{11}^{g}$.
For (12) we have $H_{12} = H_{12}^{t_0} + H_{12}^{\gr}$.

The effective Hamiltonian for the low-energy space (1) is 
\begin{align}
H_{\textrm{eff}} = (1 + H_{12} H_{22}^{-2} H_{21})^{-1}( H_{11} - H_{12} H_{22}^{-1} H_{21})
\end{align}
We  use  our decomposition of $H_{22}$ to expand
\begin{align}
H_{22}^{-1} &= \frac{1}{H_{22}^{t_1}} \sum_{n=0}^\infty  (-1)^n ( (H_{22}^{t_0} + H_{22}^{g})  \frac{1}{H_{22}^{t_1}})^{n}
\end{align} 
The leading contributions take the form
\begin{align}
H_{12}  H_{22}^{-1} H_{21} &=  H_{12}  \frac{1}{H_{22}^{t_1}}  \left[ ( H_{22}^{t_0}  \frac{1}{H_{22}^{t_1}})^{N-2} + \mathcal{O}(\frac{g}{t_1}) \right]
H_{21}   \label{eq:heffexpand}  \\
H_{12}  H_{22}^{-2} H_{21} &=  t_1^{-2} H_{12} H_{21} ( 1 + \mathcal{O}(\frac{g}{t_1})) \label{eq:renorm}
\end{align}
Let us first evaluate the result neglecting the $\mathcal{O}(\frac{g}{t_1})$ terms.
The renormalization term $(1 + H_{12} H_{22}^{-2} H_{21})^{-1}$ will multiply the rest of the Hamiltonian by terms
\begin{align}
H_{12} H_{22}^{-2} H_{21} &=  \frac{v_F \gr^{t/b}}{t_1^2}  i ( s^- k_+ - s^+ k_- )   \\
  & +  \mathcal{O}( \left( \frac{v_F k}{t_1} \right)^2, \left( \frac{\gr}{t_1} \right)^2)
\end{align}
The first term is most interesting because it breaks $s^z$, but has little influence near the $K$-point.

To leading order we can then neglect Eq.~\eqref{eq:renorm}, and writing the leading part of Eq.~\eqref{eq:heffexpand} in the $2 \times 2$ space of the bottom/top unhybridized orbital,
\begin{align}
H_{\textrm{eff}} &= \left(\begin{array}{cc} -\frac{u}{2} + \frac{m^b}{2} +   \frac{\gi^b}{2} \tau^z s^z & T  \\ T^\dagger & \frac{u}{2} -\frac{m^t}{2} + \frac{\gi^t}{2} \tau^z s^z \end{array}\right) + \frac{1}{2} \mathbf{E}^Z \cdot \vec{s} \\
T &= \frac{1}{t_1^{N-1}}(v_F k_- + \gr^b  i s^- ) (v_F k_-)^{N-2} ( v_F k_- + \gr^t  i s^- ) \\
&= \frac{ (v_F k_-)^N}{ t_1^{N-1}} + 2 (\gr^b + \gr^t) \left (\frac{v_F k_-}{t_1} \right)^{N-1} s^- 
\end{align}
The masses $m^{t/b}$ have no influence as they can be absorbed into $u$ and the chemical potential. The special case $\gr^t = -\gr^b, \gi^t = - \gi^b$ is the  case given in the text. 

Note then that for $\gr^t = -\gr^b$, $s^z$ is conserved to this order.
What is the leading term by which $\gr$ breaks $s^z$ conservation?
In powers of $\gr$, the first such term appears to be the renormalization $(1 - \frac{v_F \gr }{t_1^2}  i \sigma^z ( s^- k_+ - s^+ k_- ))$ multiplying the entire Hamiltonian, which is thus negligible, and similarly the $\mathcal{O}(\frac{g}{t_1})$ term in \eqref{eq:heffexpand} which goes as $\frac{v_F k  \gr}{t_1^2} g$.
There will also be $\mathcal{O}(\frac{\gr^2}{t_1^2} g)$ terms. 
In summary, all such terms are suppressed by factors of $t_1^{-2}$ relative to the $\mathcal{O}(g)$ terms in the Hamiltonian, and hence $s^z$ is \emph{accidentally} conserved to good approximation in this model.

Our analysis neglects the hoppings $t_2, t_3, t_4$, which contribute particle-hole asymmetry and trigonal warping.
The influence of $t_3$ is pronounced in BLG because it directly connects the unhybridized orbitals $H_{11}$,  splitting the $k^2$ dispersion into a $C_3$-symmetric configuration of $4 = |3| + |-1|$ Dirac cones.
Likewise, $t_2$ directly connects the unhybridized orbitals in TLG, splitting the $k^3$ dispersion into 3 Dirac cones.
But neither  makes a difference to our topological analysis, since sufficient $\gi$  gaps out each of the Dirac cones independently with the same net Chern number, and the TCI physics of the depends only on the representation at the $K$-point.
This is confirmed by the numerical simulations of the main text, which include these terms.

The trigonal warping does  technically change  the nature of the $\sigma_{\textrm VH} = 2$ to $\sigma_{\textrm SH} = 2$ transition in BLG. It will split into \emph{two} transitions, because there is no reason the mass of the $K$-point Dirac cone will be equal to that of the surrounding three Dirac cones, so their gap closing will happen for slightly different $u$.
The intermediate band structure is technically a strong TI with a tiny gap, but in practice we find the intermediate region is a compensated semimetal since the energies of the two types of Dirac cone are also different.

\end{document}